\documentclass[aps,twocolumn,showpacs,amsmath,amssymb]{revtex4}
\usepackage{textcomp}
\usepackage{graphicx}

\begin{document}
\title{Influence of surface tension on the conical miniscus of a
  magnetic fluid in the field of a  current-carrying wire}
\author{Thomas John, Dirk Rannacher, and Andreas Engel}
\affiliation{Institut f\"{u}r Physik, Carl von Ossietzky
  Universtit\"{a}t, 26111 Oldenburg, Germany}

\begin{abstract}
We study the influence of surface tension on the shape of the conical
miniscus built up by a magnetic fluid surrounding a current-carrying
wire. Minimization of the total energy of the system leads to a
singular second order boundary value problem for the function
$\zeta(r)$ describing the axially symmetric shape of the free surface.
An appropriate transformation regularizes the problem and allows a
straightforward numerical solution. We also study the effects a
superimposed second liquid, a nonlinear magnetization law of the
magnetic fluid, and the influence of the diameter of the wire on the
free surface profile.
\end{abstract}

\maketitle

\section{Introduction}

The shape of the free surface of a ferrofluid in a static magnetic
field is one of the prominent examples for a non-trivial interplay
between magnetic and hydrodynamic degrees of freedom in
ferrohydrodynamics: On the one hand the magnetic stresses contribute
to the force balance determining the surface profile whereas on the
other hand the local magnetic field depends on this profile due to
the magnetic boundary conditions at the surface of the magnetically
permeable material \cite{Rosensweig}. A particularily popular setup
is the conical meniscus of a ferrofluid surrounding a vertical
current-conducting wire \cite{Rosensweig}, see Fig.~1. Neglecting
surface tension the conical shape can be determined analytically
from the balance between gravitational and magnetic force
\cite{Rosensweig}.

In the present note we investigate theoretically the influence of
surface tension on the shape of the conical meniscus. Although
surface tension is known to modify free surface profiles in
ferrohydrodynamics and its influence has been studied, e.g.,  for
small gaps \cite{Flament96,Polevikov05} and capillaries
\cite{Bashtovoi05,Bashtovoi02} to our knowledge no systematic
investigation has been done so far for the conical meniscus problem.
Clearly, surface tension is present in all experiments in
ferrohydrodynamics and a full understanding of its impact on a basic
experiment in the field is hence desirable.

As expected no analytical expression for the free surface profile can
be derived when surface tension is included. The numerical
determination can be reduced to a singular boundary value problem for
an ordinary differential equation which is, however, not completely
straightforward to solve. Still, using an appropriate transformation,
accurate solutions can be obtained. The numerical effort is much
smaller than in alternative procedures such as finite difference,
Galerkin, or collocation methods which is for example used from
commercial software like ANSYS or COMSOL Multiphysics
\cite{ansys,femlab}

The paper is organized as follows. In section II we collect the basic
equations describing the system. Section III contains the analysis
of a somewhat idealized situation. In section IV we discuss the
modification of the results if the idealizations of section III are
removed. Finally section V provides some conclusions.

\section{Basic equations}

The setup to be investigated is sketched in Fig.~1. An infinitely long,
straight wire of radius $R$ oriented along the $z$-axis of a
cylindrical coordinate system $(r,\phi,z)$ carries an electric current
$I$. The wire is
surrounded by two superimposed liquids. The lower one with density
$\varrho_1$ is a ferrofluid with a given magnetization law
$\vec{M}=\vec{M}(\vec{H})$. The upper one has density $\varrho_2<\varrho_1$
and is a non-magnetic fluid. The interface between the two fluids is
characterized by an interface tension $\sigma_{1,2}$. For $I=0$ there is no
magnetic field and the flat interface between the two fluids is taken
as the $z=0$--plane of the coordinate system. For $I\neq 0$ a magnetic
field
\begin{equation}\label{H(r)}
\vec{H}(r)=\frac{I}{2\pi r}\;\vec{e_\phi}
\end{equation}
builds up which induces a force density in the
lower fluid. This gives rise to an axis-symmetric conical
interface parametrized by a function $\zeta(r)$ the determination of
which is the central aim. Note that the magnetic field (\ref{H(r)}) is
everywhere tangential to the surface and is therefore independent of
the detailed shape of the surface which makes the analysis of this case
particularily transparent.

\begin{figure}
\includegraphics[width=8.5cm]{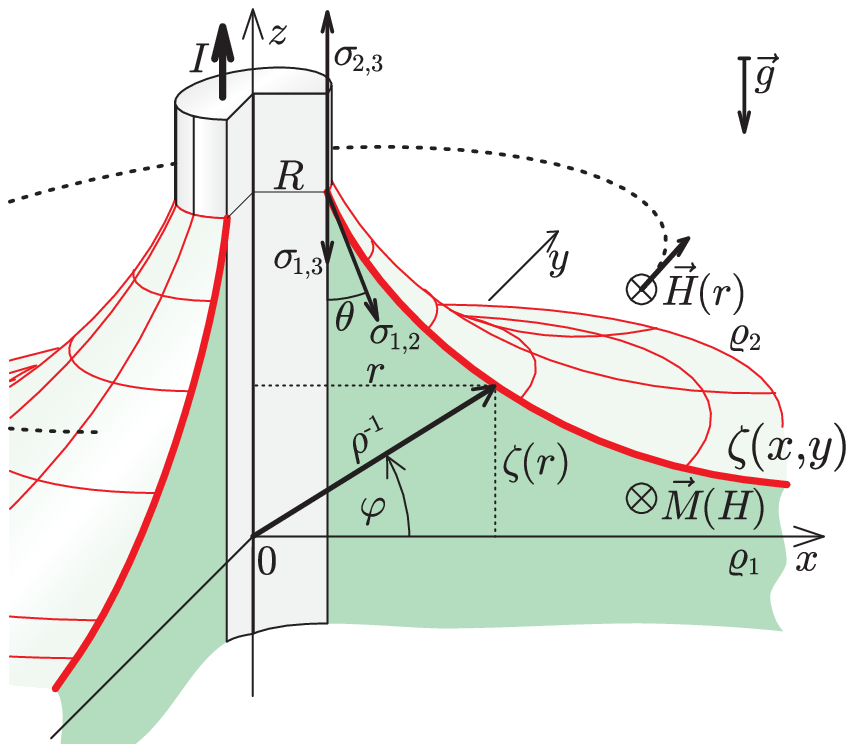}
\caption{Sketch of the setup and definition of the main
  variables. The free surface $\zeta(r)$ of the ferrofluid is
  axis-symmetric. } \label{figsetup}
\end{figure}

One way to determine $\zeta(r)$ is by minimizing the total energy
$E_\text{tot}$ of the system. It is convenient then to use the energy
of the flat interface configuration as reference state and hence to
minimize the energy difference between states with a non-trivial
$\zeta(r)$ and $\zeta\equiv 0$. Equivalently one may start with the
ferrohydrodynamic Bernoulli equation \cite{Rosensweig,Berkovski}.

The total energy difference $E_\text{tot}$ is the sum of three parts,
the gravitational, the surface, and the magnetic energy,
\begin{equation}
 E_\text{tot}= E_\text{g}+E_\text{s}+E_\text{m}. \label{Emin}
\end{equation}
The different parts are given, respectively, by
\begin{align}
 E_\text{g} & = \int_V\!\!\! \text{d}^3r\, (\varrho_1-\varrho_2) g z =
    \pi (\varrho_1-\varrho_2) g \int_R^\infty \!\!\! \text{d}r\, r\,
    \zeta(r)^2 \\
 E_\text{s} & = \int_{\partial V}\!\!\!\!\! \text{d}^2 r\,\sigma_{1,2} =
   2\pi \sigma_{1,2}\int_R^\infty \!\!\!\!\! \text{d}r\, r
   \left(\sqrt{1+\zeta'(r)^2}-1 \right)
\end{align}
and \cite{Rosensweig,LLVIII}
\begin{align}\nonumber
 E_\text{m} &= -\int_V \!\!\! \text{d}^3r\,
      \mu_0\int_0^{H(\vec{r})}\!\!\!\!
        \text{d}\vec{H}\cdot\vec{M}(\vec{H})\\
            &= -2\pi\mu_0 \int_R^\infty \!\!\!\!\text{d}r\,r \,
    \zeta(r)\int_0^{H(r)}\!\!\!\!\text{d}H M(H)  \label{E_m}
\end{align}
Here $V$ denotes that part of the volume occupied by the ferrofluid
for which $z>0$, $\partial V$ is the area of the interface between the two
fluids and $\mu_0$ is the permeability of free space. Note that in
the last equality of (\ref{E_m}) we have assumed that the
magnetization $\vec{M}$ and the magnetic field $\vec{H}$ are parallel
as is the case in ferrofluids. Note also that $H(r)$ is given by
(\ref{H(r)}).

The Euler-Lagrange equation
\begin{equation}\label{ELeq}
  \frac{\delta E_\text{tot}}{\delta \zeta(r)}=0
\end{equation}
corresponding to the minimization of
$E_\text{tot}[\zeta(r)]$ is a nonlinear ordinary differential
equation for the desired interface profile $\zeta(r)$. For a unique
solution this equation has to be complemented by appropriate boundary
conditions. These are
\begin{equation}\label{boundaryinf}
  \lim_{r\to\infty}\zeta(r)=0
\end{equation}
and
\begin{equation}\label{boundaryR}
\zeta'(R)=\frac{\sigma_{1,3}-\sigma_{2,3}}
          {\sqrt{\sigma_{1,2}^2-(\sigma_{1,3}-\sigma_{2,3})^2}},
\end{equation}
where the prime denotes differentiation with respect to $r$.
The second boundary condition results from the Young equation
\cite{Landau5}
\begin{equation}
  \sigma_{2,3}=\sigma_{1,3}+\sigma_{1,2}\cos\theta,
\end{equation}
for the contact angle $\theta$ of the fluids at the wire and the fact
that $\zeta'(R)=-\tan(\pi/2-\theta)$ (cf. Fig.~1).

\section{Simplified model}
We first analyse a somewhat simplified version of the boundary value
problem derived in the last section. The simplifications are the
following. We assume that the non-magnetic fluid is absent, $\varrho_2=0$,
that the magnetization law of the ferrofluid is linear,
$\vec{M}(H)=\chi \vec{H}$ where $\chi$ denotes the magnetic
susceptibility, and that the diameter of the wire is negligible,
$R=0$.

Since the magnetic field $H(r)$ diverges for $r\to 0$
so does the magnetic force density. In order to get a stable interface
profile the magnetic force has to be counterbalanced by gravitation
and surface tension which is possible only if
\begin{equation}\label{bc3}
  \lim_{r\to 0} \zeta(r)=\infty.
\end{equation}
This equation replaces the boundary condition (\ref{boundaryR}) in the
present case. Consequently $\sigma_{1,3}$ and $\sigma_{2,3}$ are
irrelevant and only $\sigma_{1,2}$ remains which will be denoted
simply by $\sigma$ in the present section.

Using the simplifying assumptions we find for the magnetic energy
(\ref{E_m}) the expression
\begin{equation}\label{E_mlin}
  E_\text{m} =  -\frac{\mu_0\,\chi\, I^2}{4 \pi}
               \int_0^\infty \!\!\!\!\text{d}r\,\frac{\zeta(r)}{r},
\end{equation}
which allows to explicitly perform the variation in
eq.(\ref{ELeq}). As a result we get the following differential
equation for $\zeta(r)$
\begin{equation}
  g \varrho_1 \zeta-\frac{\mu_0\chi I^2}
    {8 \pi^2}\frac{1}{r^2}-\sigma\frac{\zeta'+\zeta'^3+
    r\zeta''}{(1+\zeta'^2)^{3/2}}\frac{1}{r}=0. \label{euler}
\end{equation}
It is convenient to introduce dimensionless quantities by
measuring  both $r$ and $\zeta$ in units of
\begin{equation}
  a=\sqrt[3]{\frac{\mu_0\chi I^2}{8 \pi^2 g \varrho_1}}
\end{equation}
and $\sigma$ in units of $\varrho_1 g a^2$. Eq.~(\ref{euler}) then
acquires the form
\begin{equation}\label{dimlesseuler}
  \zeta-\frac{1}{r^2}-\sigma\frac{\zeta'+\zeta'^3 + r\zeta''}
        {(1+\zeta'^2 )^{3/2}}\frac{1}{r}=0 .
\end{equation}
This equation is easily solved for $\sigma=0$ yielding the
well-known hyperbola $\zeta(r)=1/r^2$ \cite{Rosensweig}.
A perturbative solution for $\sigma\neq 0$ by expanding $\zeta(r)$ in
powers of $\sigma$ was found to yield satisfactory results only for
values of $\sigma$ much smaller than those relevant in experiments. We
therefore turned to a numerical solution.

Eq.~(\ref{dimlesseuler}) together with the boundary conditions
(\ref{boundaryinf}) and (\ref{bc3}) represents a singular boundary
value problem of second kind (see, e.g., \cite{Asher}). The main
problem that makes a straightforward numerical solution inexpedient
are the infinite intervals for $r$ and $\zeta$. We therefore transform
both quantities according to (cf. also Fig.~1)
\begin{equation}
 \zeta(\rho,\varphi)=\frac{\sin\varphi}{\rho} \quad , \quad
  r(\rho,\varphi)=\frac{\cos\varphi}{\rho} \label{transform}
\end{equation}
and describe the interface profile by $\rho=\rho(\varphi)$. The
boundary conditions (\ref{boundaryinf}) and (\ref{bc3}) then transform
into
\begin{equation}\label{boundary2}
  \rho(\varphi=0)=0 \quad\text{and}\quad \rho\left(\varphi=\frac\pi 2\right)=0
\end{equation}
respectively which are much more convenient for the subsequent
numerical solution. Substituting the derivatives in (\ref{dimlesseuler})
according to
\begin{align}
 \zeta' &= \frac{\text{d}\zeta}{\text{d}r}=
 \frac{\rho' \sin\varphi-\rho\cos\varphi}{\rho\sin\varphi+\rho'\cos\varphi}, \\
 \zeta'' &= \frac{\text{d}^2\zeta}{\text{d}r^2}
 =-\frac{\rho^3(\rho+\rho'')}{(\rho\sin\varphi+\rho'\cos\varphi)^3},
\end{align}
we find as differential equation for $\rho(\varphi)$
\begin{widetext}
\begin{equation}\label{euler2}
\left(\sin\varphi-\rho^3\sec^2\varphi\right)
    \left(\rho^2+\rho'^2\right)^{3/2}
+\sigma\rho^2\left(\rho\rho'^2-\rho'^3\tan\varphi-\rho^2\rho'\tan\varphi
+\rho^2\left(2\rho+\rho''\right)\right)=0.
\end{equation}
\end{widetext} Rewriting this equation in the form
$\rho''(\varphi)=f(\varphi,\rho(\varphi),\rho'(\varphi))$ the
boundary value problem (\ref{euler2}), (\ref{boundary2}) can now
easily be solved numerically using, e.~g., the nonlinear
finite-difference method described in \cite{Faires}. As initial guess
for the solution which is needed in this procedure the transformation
of the analytical solution $\zeta_0(r)=1/r^2$ for $\sigma=0$ given by
$\rho_0(\varphi)=\cos(\varphi)\sqrt[3]{\tan(\varphi)}$ may be used.

\begin{figure}
\begin{center}
\includegraphics[width=8.5cm]{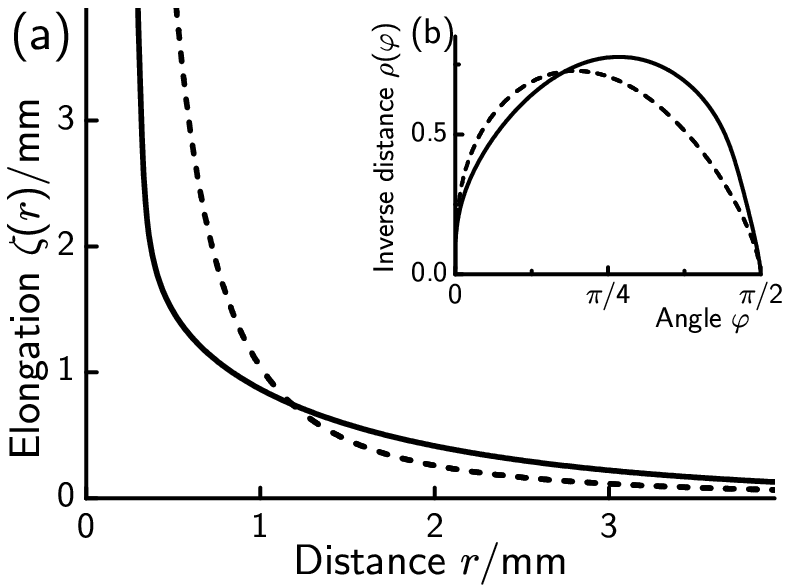}
\end{center}
\caption{(a) Free surface profile $\zeta(r)$ as determined numerically
  for $R=0$ and $I=30\,$A for the ferrofluid EMG909 with parameters
  $\varrho_1=1120$\,kg\,m$^{-3}$, $\sigma=0.0259$\,N\,m$^{-1}$, and
  $\chi=0.8$ (full line). Also shown is the result for $\sigma=0$ and
  otherwise identical parameters (dashed line). (b) Same profiles as
  in (a) in terms of the transformed function $\rho(\varphi)$.}
\label{figrequal0}
\end{figure}

Fig.\ref{figrequal0} shows the resulting interface profile for
parameter values corresponding to the ferrofluid EMG909 of Ferrotec
\cite{Ferrotec}. For comparison the shape for $\sigma=0$ is also
shown. It is clearly seen that the free surface profile is markedly
modified by the influence of the surface tension. As expected the
inclusion of surface tension makes the profile narrower since an
additional force pointing radially inward builds up.

\section{More realistic situations}

The above solution of the idealized problem forms a convenient starting
point for the discussion of the influence of those feature in the
original setup that were neglected in the previous section.

\begin{figure*}[th]
\begin{center}
\includegraphics[width=\textwidth]{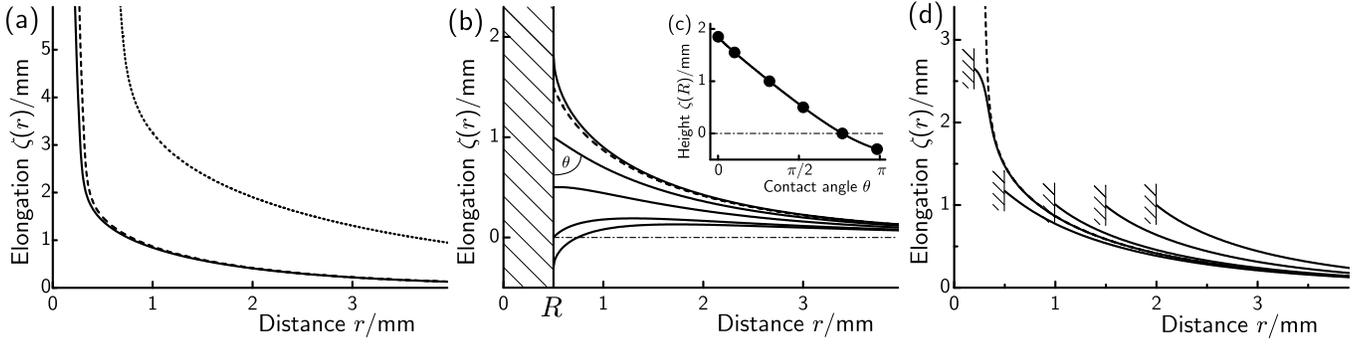}
\end{center}
\caption{Modifications of the idealized results of section 3. If not
  explicitly stated otherwise the parameters are the same as in
  Fig.~\ref{figrequal0}. The dashed line shows always the same
  interface profile as the full line in Fig.~\ref{figrequal0}.
  (a) The dotted line shows the interface profile if a non-magnetic
  fluid with density  $\varrho_2=10^3\,\text{kg\,m}^{-3}$ is
  superimposed to the ferrofluid. Also in (a), the solid line shows
  the interface profile for a ferrofluid with magnetization law of
  Langevin type with $M_\text{s}=16$\,kA\,m$^{-1}$ and the same
  zero-field susceptibility $\chi$ as in Fig.~\ref{figrequal0}.
  (b) Dependence of the interface profile on the contact angle
  $\theta$ for non-zero radius $R=0.5$\,mm. (c) Relation between
  the contact angle $\theta$ and the height of the interface at the
  wire $\zeta(R)$. Marked points correspond to the profiles shown in
  (b). (d) Sequence of surface profiles with fixed contact angle
  $\theta=45$\textdegree~ for decreasing diameter of the wire,
  $R=2;1.5;1;0.5;0.2$\,mm. For $R\to 0$ the profile of
  Fig.~\ref{figrequal0} is almost reproduced. \label{figcorrections}}
\end{figure*}

A simple modification is the case in which the ferrofluid is superimposed by a
non-magnetic liquid. Then the density $\varrho_1$ has to substituted by the
density difference $\varrho_1-\varrho_2$ and the surface tension $\sigma$ is to
be replaced by the interface tension $\sigma_{1,2}$. The gravitational
contribution to the total energy gets reduced and consequently larger
displacements from the flat interface are to be expected. The situation can be
analyzed quantitatively without further effort by mapping it to the case
analyzed in the previous section. In fact the interface profile is again
determined by eq.~(\ref{dimlesseuler}) with only the dimensionless units and
the value of $\sigma$ being modified. As an explicit example we show in
Fig.\,\ref{figcorrections}(a) how the interface profile of
Fig.~\ref{figrequal0} gets modified for $\varrho_2=10^3$\,kg\,m$^{-3}$
and $\sigma_{1,2}=\sigma$.

Deviations form the linear magnetization law which become relevant in
particular for small values of $r$ where the magnetic field gets strong can
also be dealt with. An improved approximation is provided by the Langevin
relation \cite{Rosensweig},
\begin{equation}
M(H)=M_\text{s}\left(\coth(\alpha)-\frac{1}{\alpha}\right) \quad ,
\quad \alpha=\frac{m H}{k_\text{B}T},
\end{equation}
with the saturation magnetization of the fluid $M_\text{S}$, the
magnetic moment $m$ of a single ferromagnetic particle, the
Boltzmann-constant $k_\text{B}$ and the temperature $T$. The
zero-field susceptibility is then given by $\chi= m
M_\text{S}/(3k_\text{B}T)$. The density of the magnetic energy can
again be determined analytically
\begin{equation}
  \int_0^{H(\vec{r})}\!\!\!\!\text{d}\vec{H}\cdot\vec{M}(\vec{H})=
   \frac{M_\text{S}k_\text{B}T}{m} \ln\frac{\sinh\alpha}{\alpha}  .
\end{equation}
This expression gives rise to a somewhat more complicated equation
for $\zeta(r)$, however, the overall procedure of section III remains
valid. As shown in Fig.~\ref{figcorrections}(a) the modification of
the result for linear magnetization law is usually small in
accordance with the fact that the fraction of the ferrofluid volume
for which the field is large  is rather small.

Finally, to elucidate the influence of a non-zero radius of the wire
we consider the problem for $R>0$ but with $\varrho_2=0$ and a linear
magnetization law. The main qualitative difference is that instead of
(\ref{bc3}) we have to use (\ref{boundaryR}) as boundary condition
for small $r$. This implies that $\zeta(r)$ is not singular at the
lower boundary. Hence already the simple transformation $r=1/u$ is
sufficient to get a numerically tractable boundary value problem for
the function $\zeta(u)$. As initial guess a linear dependence
$\zeta_0(u)=\lambda u$ may be used.

Note that $\lambda$ fixes the value of $\zeta(r)$ at $r=R$ whereas the
boundary condition (\ref{boundaryR}) involves the derivative
$\zeta'(r)$ at $r=R$. Therefore $\lambda$ or equivalently $\zeta(R)$
has to be modified until the required value for $\zeta'(R)$ is
obtained. This procedure
leads to a monotonic relation between the prescribed contact angle
$\theta$ and $\zeta(R)$, cf. Fig.~\ref{figcorrections}(c). With the
help this value for $\zeta(R)$ the complete surface profile
$\zeta(r)$ can then be determined. Fig.~\ref{figcorrections}(b) shows
a collection of surface profiles with different $\zeta(R)$ and
corresoponding contact angles $\theta$. In the limiting case
$R\to 0$, the surface profile is independent of the contact
angle $\theta$ as demonstrated in Fig.~\ref{figcorrections}(d)
for a sequence of profiles with different values of $R$ and a fixed
contact angle $\theta=45$\textdegree.

\section{Conclusions}

In the present investigation we have determined the free surface
profile of a magnetic fluid surrounding a vertical current carrying
wire with special emphasis on the effects of interface and surface
tension respectively. We have found that for experimentally relevant
parameter values there is a strong influence of the surface tension
giving rise to a more slender profile as compared to the well-studied
case without surface tension \cite{Rosensweig}. A more general
magnetization law including saturation as well as the prescription of
the contact angle at the wire changes the profile only in the vicinity
of the wire. A superimposed non-magnetic liquid changes the overall
scale of the profile due to the reduction of the gravitational
energy.

\bibliographystyle{elsart-num}

\end{document}